# Tails of Localized Density of States of Two-dimensional Dirac Fermions

Simon Villain-Guillot[a,b]

[a]*Max-Planck-Institut für Physik Komplexer Systeme, Nöthnitzer Str. 38, D-01187 Dresden, Germany*
[b]*Centre de Physique Théorique et de Modélisation, Université Bordeaux I,
CNRS-ERS 2120, F-33405 Talence Cedex, France*

Giancarlo Jug[a,c] and Klaus Ziegler[a,d]

[c]*INFM - UdR Milano, and Dipartimento di Scienze Matematiche, Fisiche e Chimiche,
Università dell'Insubria, Via Lucini 3, I-22100 Como, Italy,
and INFN – Sezione di Pavia, Italy*
[d]*Institut für Physik, Universität Augsburg, D-86135 Augsburg, Germany*

(November 15, 2018)

The density of states of Dirac fermions with a random mass on a two-dimensional lattice is considered. We give the explicit asymptotic form of the single-electron density of states as a function of both energy and (average) Dirac mass, in the regime where all states are localized. We make use of a weak-disorder expansion in the parameter $g/m^2$, where $g$ is the strength of disorder and $m$ the average Dirac mass for the case in which the evaluation of the (supersymmetric) integrals corresponds to non-uniform solutions of the saddle point equation. The resulting density of states has tails which deviate from the typical pure Gaussian form by an analytic prefactor.

71.23.An, 72.15.Rn, 73.20.Dx

## I. INTRODUCTION

Dirac fermions in two dimensions play a prominent role in the description of some notable two-dimensional (2D) condensed matter systems. Examples are the plateau transition in the Integer Quantum Hall Effect (IQHE) [1,2], $D$–wave superconductivity in $CuO$-planes [3] and quasiparticles in the resonant valence bond state of the two-dimensional Heisenberg model [4]. Notwithstanding, of course, the 2D Ising model in the framework of a Grassmann field representation (for a review, see [5]). The Dirac mass creates a gap between the hole and the particle band and, therefore, gives an opportunity for the description of an insulating fermionic system when the Fermi energy lies inside this gap. The tuning of the mass allows us to go through a transition to a metallic state when the mass (i.e. the gap) vanishes. This behavior is relevant for the description of, e.g., the plateau transition in the IQHE. However, at zero energy (measured with respect to the Fermi level) the density of states (DOS) is always zero, as it turns out, for the pure Dirac fermions, even for the case of a vanishing mass.

In practice, in a real system there are always relevant impurities such that the pure Dirac fermions model is insufficient as a description. In order to take these impurities into accout we can simply add randomness to the Dirac fermions' Hamiltonian. This has severe consequences on the properties of the Dirac fermions; in particular, as it turns out, randomness is a non-perturbative effect [6]. Randomness can lead to new (localized) states even in the gap of the pure system due to the formation of Lifshitz tails. This phenomenon will be studied in the present paper for the limit case of a large mass. We stress that, although there are different types of randomness which can be in principle included in the Dirac model, we will concentrate here on the physically relevant case of a random Dirac mass only. Moreover, we will only investigate here the form of the average DOS for the localised states. The average DOS has been evaluated via various other approaches. Fradkin, for example, computed this quantity for Dirac fermions with a random energy [7] by means of the coherent potential approximation (CPA) and found a filling of the pure system's gap. In the case of a Lorentzian mass distribution, e.g. $P(M_r) = (\tau^2 + (M_r - m)^2)^{-1}\tau/\pi$, the average DOS can be even calculated exactly [8]. At energy $E = 0$ it has, in contrast to the vanishing DOS of the pure Dirac fermions's case, a non-zero value

$$\rho(m, E = 0) \equiv \rho(m) = \tau \log[1 + \lambda/(\tau^2 + m^2)]. \tag{1.1}$$

The lattice constant $1/\lambda$ shows up as a cut-off for an infrared divergency. The power law of the tails at large mass $m$ is an artefact of the Lorentzian distribution, which has strong tails itself. For more realistic distributions, like the Gaussian $P(M_r) = \exp(-(M_r - m)^2/2g)/\sqrt{\pi g}$, we expect much weaker tails; for this case the renormalization group (RG) calculation gives a vanishing DOS. This is a consequence of the fact that a random mass is an infrared marginally-irrelevant perturbation which drives the system always back to free Dirac fermions' theory as its fixed point [9,3,10,11]. This somewhat bizarre result must be traced back to the intrinsic perturbative nature of the RG calculation: the $\beta$-function is indeed always evaluated by means of a perturbative expansion of the disorder strength $g$. Although the solution of the RG equations may pick up some non-perturbative feature for small $g$, since these equations make statements valid to all orders, there is nevertheless no guarantee that *all* non-perturbative effects are duly taken into account. There are in fact rigorous estimates of the DOS at $m = 0$ which give a deviation from the vanishing DOS of the free Dirac fermions' fixed point [6]

$$\rho(0) > c_1 e^{-c_2/g}, \tag{1.2}$$

where $c_1 > 0$ and $c_2 > 0$ are constants independent of $g$. This result is similar to the one from the (uniform) saddle-point (SP) approximation (large $N$ limit), where $\rho(0) \sim e^{-\pi/g}/g$ was found [12].

For the Dirac fermions, a non-perturbative approach based on the SP approximation of a suitable functional-integral representation has given, again for the extended states, a finite bandwidth in the DOS having a characteristic semi-circular form [12]. Perturbative calculations (leading, however, to different and sometime unphysical results) have also concentrated on the extended states. Therefore, hitherto no specific information on the features of the DOS for the *localized* states has been obtained from the Dirac fermion approach and it certainly of some great interest to characterize the localized states in view of a possible resolution of the still open problem of the localization length near the Integer Quantum Hall Transition (IQHT).

In this paper, we produce such a calculation for the Dirac fermions model characterized by a Gaussian random mass distribution. In order to make the calculation feasible, we consider the behavior of the non-uniform SP solution in the small $g/m^2$ regime. The behavior for small $g/m^2$ can be in fact explained in terms of an SP approach to the supersymmetric functional integral for the averaged DOS [13,14]. Instead of doing the SP integration, we choose to expand the functional integral directly in powers of $1/m$.

## II. THE MODEL

In this Note we carry out an investigation on the form of the localized DOS at zero energy, which leads us to characterizing the tails of this quantity in the asymptotic limit of a large average mass (or, as it turns out, in the limit of weak disorder in the single-particle random potential). Our starting point is the Dirac fermions' Hamiltonian for the independent quasiparticles

$$H_D = i\nabla_1 \sigma_1 + i\nabla_2 \sigma_2 + M\sigma_3 \equiv i\nabla \cdot \sigma + M\sigma_3, \tag{2.1}$$

where the energy is measured in units of the hopping parameter of some original lattice model, $\nabla_j$ is the lattice differential operator in the $j$-direction and $\{\sigma_j\}$ are the Pauli matrices. This Hamiltonian, with a random mass term $M_r$ and no random vector potential, is a reasonable starting point for the description of a number of interesting physical systems, as discussed in the Introduction. The average local DOS is obtained as usual from the averaged one-particle Green function $G(E - i\omega) = (H_D - E + i\omega)^{-1}$ as

$$\rho(m, E) = -\lim_{\omega \to 0} \frac{1}{\pi} Im \langle Tr_2 G_{r,r}(M, E - i\omega) \rangle, \tag{2.2}$$

where $Tr_2$ stands for the trace over the $2 \times 2$ matrix structure. We have characterized the random Dirac mass effectively by its average value $\langle M_r \rangle = m$ and correlation function $\langle M_r M_{r'} \rangle = g\delta_{r,r'}$. The representation for the Green's function in terms of both commuting and anticommuting functional integrals is well-known [12] and we arrive at a collective fields representation which for the Green's function with $r' = r$ gives

$$\langle G_{r,r}(M, E - i\omega) \rangle = \frac{1}{g} \int \tau Q_r \tau e^{-S} \mathcal{D}P \mathcal{D}Q \mathcal{D}\Theta \mathcal{D}\bar{\Theta}. \tag{2.3}$$

$\tau$ is here the diagonal matrix $(1, i)$, i.e $\tau^2 = \sigma_3$. The mixed commuting-anticommuting (supersymmetric) effective action has the form (setting, for convenience, $E = 0$)

$$\begin{aligned}S = &\frac{1}{g}\sum_r (Tr_2 Q_r^2 + Tr_2 P_r^2 + 2Tr_2 \Theta_r \bar{\Theta}_r) + \\ &+ \ln\det[(H_0 + i\omega\sigma_0 - 2\tau Q\tau)(H_0^T + i\omega\sigma_0 + 2i\tau P\tau)^{-1}] \\ &+ \ln\det[\mathbf{1} - 4\tau\bar{\Theta}\tau(H_0^T + i\omega\sigma_0 + 2i\tau P\tau)^{-1}\tau\Theta\tau(H_0 + i\omega\sigma_0 - 2\tau Q\tau)^{-1}].\end{aligned} \tag{2.4}$$

Here, $H_0$ is the Dirac Hamiltonian without disorder : $H_0 = i\sigma \cdot \nabla + m\sigma_3$.

# III. SADDLE POINT APPROXIMATION

At $E = 0$ the DOS of the pure Dirac fermions is always vanishing. Near $m = 0$ – the point where the pure system closes the energy gap – the above supersymmetric effective action enables us to recover the average DOS of the random system. From the uniform solution for the SP equation ($\delta S = 0$) in the matrix argument $Q_r$ we obtain a semicircular DOS with radius $m_c = 2e^{-\pi/g}$ with respect to its dependence on the mass $m$ [12]. Thus this predicts, within a large-$N$ approach, a finite bandwidth for the states between two critical points $m = \pm m_c$.

In order to work with localized states, a *non-uniform* solution for the SP equation for $Q_r$ must be sought, as in the approach by Cardy [13] and Brézin [14], who considered a random Schrödinger Hamiltonian using the bosonic replica trick. The situation is more complicated in the case of our supersymmetric field theory because of the presence of anticommuting variables. To avoid the difficulty of solving a differential equation with both commuting and anticommuting field components, only a schematic discussion in terms of the classical field equation (SP equation) is given in this section. From these qualitative arguments we then devise a procedure to obtain the large $m$ behavior of the field theory.

In order to obtain a non-uniform (soliton-like) solution for the SP equation of our action $S$, Eq. (2.4), it is convenient to work in the continuum limit of this model. For this purpose we introduce the characteristic length scale $a/\sqrt{g}$ on the $\sum_r$. Then the continuum limit, $a/\sqrt{g} \to 0$, can be taken. Our action thus transforms as follows

$$S = \frac{a^2}{g} \sum_r (\mathrm{Tr}_2 Q_r^2 + \cdots) \to \int (\mathrm{Tr}_2 Q_r^2 + \cdots) d^2r. \tag{3.1}$$

The SP method leads to the equation

$$\sigma_3 \tau Q_r \tau \sigma_3 = g(H_0 + i\omega\sigma_0 - 2\tau Q\tau)_{rr}^{-1}. \tag{3.2}$$

We notice that we can replace $H_0$ by $m\sigma_3 + i\sigma \cdot \nabla$. Then the SP equation can be expanded in powers of $1/m$ because we are only interested in the large $m$ behavior. To this end we pull out a factor $m\sigma_3$

$$\sigma_3 \tau Q_r \tau \sigma_3 = g\frac{\sigma_3}{m}(\mathbf{1} + \frac{i}{m}\sigma \cdot \nabla \sigma_3 + i\frac{\omega}{m}\sigma_3 - \frac{2}{m}\tau Q \tau \sigma_3)_{rr}^{-1} \tag{3.3}$$

and multiply by $\sigma_3$ from both sides, which gives

$$\tau Q_r \tau = \frac{g}{m}(\mathbf{1} + \frac{i}{m}\sigma \cdot \nabla \sigma_3 + i\frac{\omega}{m}\sigma_3 - \frac{2}{m}\tau Q \tau \sigma_3)_{rr}^{-1}\sigma_3. \tag{3.4}$$

Using the notation $\bar{Q}_r = \tau Q_r \tau$ and taking the limit $\omega = 0$, the SP equation reads, in leading order in powers of $1/m$ and in lowest non-trivial order in $\nabla Q_r$

$$\bar{Q}_r - \frac{g}{m}\sigma_3 = \frac{g}{m}\Big[\frac{2}{m}\bar{Q}_r + \frac{4}{m^2}\bar{Q}_r \sigma_3 \bar{Q}_r - \frac{2}{m^3}(\sigma \cdot \nabla \sigma_3 \bar{Q} \sigma_3 \sigma \cdot \nabla)_{rr} + \cdots\Big]. \tag{3.5}$$

In the expansion we have taken a second order term in $\bar{Q}$ as well as the second order gradient term (the first order term being traceless). The constant term can be removed by introducing the shifted new field

$$\tilde{Q}_r = \bar{Q}_r - \frac{g}{m}\sigma_3$$

and by keeping only the leading order in $1/m$ for each power of $\tilde{Q}_r$. Then the SP equation can be written schematically (omitting the Pauli matrices) as

$$\frac{1}{g}\tilde{Q}_r = \frac{a}{m^3}\tilde{Q}_r^2 + \frac{b}{m^4}(\nabla^2 \tilde{Q})_r, \tag{3.6}$$

where $a$ and $b$ are numbers. This equation can be interpreted as the SP equation for the following effective action

$$S_{eff} = \int \Big(\frac{1}{2}\tilde{Q}_r^2 - g\frac{a}{3m^3}\tilde{Q}_r^3 - g\frac{b}{2m^4}(\nabla \tilde{Q})_r^2\Big)d^2r. \tag{3.7}$$

We are interested in finding the dependence of the effective action on the physical parameters $m$ and $g$. For this purpose we use as an ansatz for the non-linear SP equation, Eq. (3.6)

$$\tilde{Q}_r = \frac{m^3}{g}\mathcal{Q}(\frac{m^2}{\sqrt{g}}r),$$

where $\mathcal{Q}$ is now some dimensionless function. Moreover, we introduce the rescaled two-dimensional space variable

$$R = \frac{m^2}{\sqrt{g}}r.$$

Then we obtain from the differential equation (3.6) a non-linear differential equation for $\mathcal{Q}(R)$, where the coefficients are numbers

$$\frac{m^3}{g^2}(\mathcal{Q} - a\mathcal{Q}^2 - b\nabla_R^2\mathcal{Q}) = 0. \tag{3.8}$$

Furthermore, the effective action reads in terms of $\mathcal{Q}(R)$ and

$$\frac{m^2}{g}\int[\frac{1}{2}\mathcal{Q}^2 - \frac{a}{3}\mathcal{Q}^3 - \frac{b}{2}(\nabla_R\mathcal{Q})^2]d^2R.$$

Since the integral is now just a number (it depends, of course, on the function $\mathcal{Q}(R)$ which needs to be determined as a solution of the differential equation (3.8)), the action is proportional to $m^2/g$. This implies that for large $m$ the DOS must be of the form

$$\rho(m) \sim \rho_0 e^{-\alpha m^2/g}.$$

The parameter $\alpha$ and the coefficient $\rho_0$ of the exponential factor will be evaluated below.

## IV. BEHAVIOR FOR A LARGE DIRAC MASS

Having determined the functional dependence for the DOS via a saddle point approximation, in the following we shall use a more direct route by focusing on the large-$m$ limit, or on the tail corrections, without actually using the saddle point approximation. The approach we use is based on an expansion of the action $S$ in powers of $1/m$ up to lowest order in this parameter. It will result in an expansion in the parameter $g/m^2$, so the following calculation holds good for the tails of the localized states in the DOS, and for weak disorder.

For this purpose, we can separate the off-diagonal contribution of $H_0$, namely $H_0' = i\sigma \cdot \nabla$, from the logarithmic term of the action of Eq. (2.4), as follows:

$$H_0 + i\omega\sigma_0 - 2\tau Q\tau = B_q + H_0'$$
$$H_0 + i\omega\sigma_0 + 2i\tau P\tau = B_p + H_0'. \tag{4.1}$$

Then we expand the logarithmic term of the action around the diagonal part of the Hamiltonian. We obtain, e.g., for the $Q$-dependent expression to leading order in the off-diagonal term

$$\ln\det(H_0 + i\omega\sigma_0 - 2\tau Q\tau) = \ln\det(m\sigma_3 + i\omega\sigma_0 - 2\tau Q\tau) + \text{Tr}(H_0'B_q^{-1}). \tag{4.2}$$

Thus, for the complete expression we get, from the expansion in terms of the off-diagonal part $H_0'$

$$e^{-S} = e^{-S'}[1 - \text{Tr}(H_0'B_q^{-1}) + \text{Tr}(H_0'B_p^{-1}) + 4\text{Tr}(\tau\bar{\Theta}\tau B_p^{-1}\tau\Theta\tau B_q^{-1}) + \cdots] \tag{4.3}$$

with the unperturbed part of the action

$$S' = \frac{1}{g}\sum_r(\text{Tr}_2 Q_r^2 + \text{Tr}_2 P_r^2 + 2\text{Tr}_2\Theta_r\bar{\Theta}_r) - \ln\det(B_q) + \ln\det(B_p). \tag{4.4}$$

This is a local action and all non-local expressions appear in the expansion terms. This means that the unperturbed "weight" $e^{-S'}$ factorizes on the lattice. In particular, we obtain for the diagonal part of the Green's function in leading order the expression

$$\int Q_{r_0} \frac{\prod_r \det_2(m\sigma_3 + i\omega\sigma_0 + i2\tau P_r\tau)}{\prod_r \det_2(m\sigma_3 + i\omega\sigma_0 - 2\tau Q_r\tau)} e^{-S''} \mathcal{D}P_r \mathcal{D}Q_r \mathcal{D}\Theta_r \mathcal{D}\bar{\Theta}_r$$
$$= \int Q_{r_0} \mathcal{P}(r_0) \mathcal{D}P_{r_0} \mathcal{D}Q_{r_0} \mathcal{D}\Theta_{r_0} \mathcal{D}\bar{\Theta}_{r_0} \prod_{r \neq r_0} \int \mathcal{P}(r) \mathcal{D}P_r \mathcal{D}Q_r \mathcal{D}\Theta_r \mathcal{D}\bar{\Theta}_r, \tag{4.5}$$

where $S'' = \frac{1}{g} \sum_r (\text{Tr}_2 Q_r^2 + \text{Tr}_2 P_r^2 + 2\text{Tr}_2 \Theta_r \bar{\Theta}_r)$ and $\mathcal{P}(r)$ is the $r$-dependent part of the integrand (i.e. the ratio of the two determinants times the corresponding factor of $e^{-S''}$). Due to the supersymmetric representation and because terms involving $B_{p_r}$ and $B_{q_r}$ are local operator at site $r$, we have the following relation [15]:

$$\int \mathcal{P}(r) \mathcal{D}P_r \mathcal{D}Q_r \mathcal{D}\Theta_r \mathcal{D}\bar{\Theta}_r = 1. \tag{4.6}$$

Thus we are left with the computation of the integral at the site $r_0$, namely:

$$\int Q_{r_0} \frac{\det_2(m\sigma_3 + i\omega\sigma_0 + i2\tau P_{r_0}\tau)}{\det_2(m\sigma_3 + i\omega\sigma_0 - 2\tau Q_{r_0}\tau)}$$
$$\exp[-\frac{1}{g}(\text{Tr}_2 Q_{r_0}^2 + \text{Tr}_2 P_{r_0}^2 + 2\text{Tr}_2 \Theta_{r_0} \bar{\Theta}_{r_0})] \mathcal{D}P_{r_0} \mathcal{D}Q_{r_0} \mathcal{D}\Theta_{r_0} \mathcal{D}\bar{\Theta}_{r_0}. \tag{4.7}$$

For the Hermitean matrices $Q_{r_0}$, $P_{r_0}$ we use the following parametrizations

$$Q_{r_0} = \begin{pmatrix} q_0 + q_3 & q_1 - iq_2 \\ q_1 + iq_2 & q_0 - q_3 \end{pmatrix}, \quad P_{r_0} = \begin{pmatrix} p_0 + p_3 & p_1 - ip_2 \\ p_1 + ip_2 & p_0 - p_3 \end{pmatrix}.$$

This gives, for instance

$$\text{Tr}_2 Q_{r_0}^2 + \text{Tr}_2 P_{r_0}^2 = 2(p_0^2 + p_2^2 + p_2^2 + p_3^2 + q_0^2 + q_1^2 + q_2^2 + q_3^2).$$

We can immediately perform the $P_{r_0}$-integration (see Appendix A), obtaining

$$\int \det_2(m\sigma_3 + i\omega\sigma_0 + i2\tau P_{r_0}\tau) \exp(-\frac{1}{g}(\text{Tr}_2 P_{r_0}^2)) \mathcal{D}P_{r_0} = -(\frac{\pi}{2}g)^2 (\omega^2 + m^2 + 2g). \tag{4.8}$$

The integration over the Grassmannian fields $\Theta$ and $\bar{\Theta}$ contributes a factor $(\frac{2}{\pi g})^4$. Then the DOS reads, according to Eqs. (2.2) and (2.3)

$$\rho(m) = \frac{4}{\pi^3 g^3} \lim_{\omega \to 0} (\omega^2 + m^2 + 2g) \mathcal{I}m \int \frac{q_3}{\det_2(B_q)} e^{-2(q_0^2 + q_1^2 + q_2^2 + q_3^2)/g} dq_0 \cdots dq_3. \tag{4.9}$$

At this point, we have to carry out the integration with respect to $Q_{r_0}$. This integral is evaluated in Appendix B and gives, finally

$$\rho(m) = \frac{1}{8\sqrt{\pi g}} (1 + m^2/2g)^2 e^{-m^2/4g}. \tag{4.10}$$

The typical form of the DOS, as a function of $m$ and at the Fermi energy (i.e. $E = 0$) is shown in Figure 1 (where a comparison with the corresponding extended-states (uniform SP solution) semi-circular DOS is also shown). We can now see that the DOS becomes, in the limit of large $m$ and as anticipated, a function of $m^2/g$ which characterizes the tails of the localized electron states distribution. These tails become more pronounced as the disorder, and thus $g$, increases. The exponential dependence on $m^2/g$ is characteristic of a soliton-like solution of the SP approximation. In particular, we are now able to determine the parameters of the SP approximation of the previous Section as

$$\alpha = 1/4, \quad \rho_0 = \frac{1}{8\sqrt{\pi g}} (1 + m^2/2g)^2$$

without solving the SP equation (3.8) directly.

## V. CONCLUSIONS

Dirac fermions have two bands with an energy gap proportional to the mass $m$. We evaluate the DOS at $E = 0$, i.e. in the middle of the gap, whilst varying the mass $m$. This means that we measure the DOS of the tails which develop inside the gap and find that the DOS acquires the asymptotic form $(1 + m^2/2g)^2 e^{-m^2/4g}$ for small values of $g/m^2$. Thus the behavior is controlled by the the exponential part $e^{-m^2/4g}$. Conversely, in the case of ordinary Schrödinger particles subject to a Gaussian white-noise potential, Cardy [13] and (with a different prefactor) Brèzin and Parisi [14] found tails in the energy $E$ for the DOS which follow the more simple exponential decay. In our case, the Gaussian decay of the DOS with $m$ reflects the Gaussian distribution of the Dirac mass. With other distributions for the disorder, we would expect also other forms for the DOS tails.

We have addressed the question of the existence of localized states created by a random mass in the gap of the Dirac fermions' model in 2D. Our findings are in reasonable agreement with previous calculations for this model based on the homogenous SP method [8] and for 2D Dirac fermions with a random energy term [7]. It is interesting that the expansion parameter is here $g/m^2$ (i.e. that the expansion is valid for weak randomness $g$), whereas localized states are usually related to strong randomness. Of course, it is crucial that also a relatively large mass $m$ is present in order to obtain a valid expansion parameter. This regime is complementary to what we have in the homogeneous saddle point approximation where weak randomness is considered at any mass $m$. In the latter we have critical points $m = \pm m_c \equiv \pm 2 e^{-\pi/g}$. It would be interesting to approach these critical points, which are related to mobility edges with extended states for small $|m|$, using the non-local method developed in this paper. This could contribute, e.g., to the long-standing problem of the divergence of the localization length $\xi(m)$ as the mobility edge is approached.

### Acknowledgements

One of us (GJ) is grateful to Prof. K.B. Efetov and to the DFG via Sonderforschungsbereich 237 (Unordnung und grosse Fluktuationen, Projekt A8: "Fluktuationen in mesoskopischen und ungeordneten Systeme") at the Ruhr-Universität Bochum, for partial support during the initial stage of this work. He is also grateful to Prof. P. Fulde and to the Max-Planck-Gesellschaft for support at the MPI für Physik Komplexer Systeme in Dresden.

### APPENDIX A

In this Appendix we work out the $P$-block integration. With the expression

$$B_p = m\sigma_3 + i\omega\sigma_0 + 2i\tau P\tau = \begin{pmatrix} i\omega + 2ip_3 + m + 2ip_0 & 2ip_2 - 2p_1 \\ -2ip_2 - 2p_1 & i\omega + 2ip_3 - m - 2ip_0 \end{pmatrix}, \quad (A.1)$$

so that

$$-\det{}_2 B_p(r_0) = (\omega + 2p_{3,r_0})^2 + (m + 2ip_{0,r_0})^2 + 4p_{2,r_0}^2 + 4p_{1,r_0}^2, \quad (A.2)$$

we see that the integration over the 2×2 $P$-block is easily carried out. Bearing in mind that $\text{Tr}_2(P_{r_0}^2) = 2\sum_i p_{i,r_0}^2$, we get

$$-\int dp_1 dp_2 \det{}_2 B_p e^{-\frac{2}{g}(p_1^2 + p_2^2)} = \frac{\pi}{2}g^2 + \frac{\pi}{2}g[(\omega + 2p_3)^2 + (m + 2ip_0)^2], \quad (A.3)$$

$$\int dp_3 [\frac{\pi}{2}g^2 + \frac{\pi}{2}g((\omega + 2p_3)^2 + (m + 2ip_0)^2)] e^{-\frac{2}{g}p_3^2}$$
$$= [\frac{\pi}{2}g^2 + \frac{\pi}{2}g(\omega^2 + (m + 2ip_0)^2)](\frac{\pi}{2}g)^{\frac{1}{2}} + \pi^{\frac{3}{2}}(\frac{g}{2})^{\frac{5}{2}} \quad (A.4)$$

and finally

$$\int dp_0 \left\{ [\frac{\pi}{2}g^2 + \frac{\pi}{2}g(\omega^2 + (m + 2ip_0)^2)](\frac{\pi}{2}g)^{\frac{1}{2}} + (\frac{g}{2})^{\frac{5}{2}}\pi^{\frac{3}{2}} \right\} e^{-\frac{2}{g}p_0^2} = (\frac{\pi}{2}g)^2 [\omega^2 + m^2 + 2g]. \quad (A.5)$$

## APPENDIX B

At this point we carry out the integration over the elements of the $Q$-block. We first of all notice that

$$B_q = m\sigma_3 + i\omega\sigma_0 - 2\tau Q\tau = \begin{pmatrix} i\omega - 2q_3 + m - 2q_0 & -2q_2 - i2q_1 \\ 2q_2 - i2q_1 & i\omega - 2q_3 - m + 2q_0 \end{pmatrix} \quad \text{(B.1)}$$

so that $\det_2 B_q = (i\omega - 2q_3)^2 - (m - 2q_0)^2 + 4q_2^2 + 4q_1^2$. Then we can write

$$\frac{4}{\det_2 B_q} = \frac{q_1^2 + q_2^2 + q_3^2 - (q_0 - \frac{m}{2})^2 - \frac{\omega^2}{4} + i\omega q_3}{(q_1^2 + q_2^2 + q_3^2 - (q_0 - \frac{m}{2})^2 - \frac{\omega^2}{4})^2 + \omega^2 q_3^2}. \quad \text{(B.2)}$$

The contribution to the imaginary part of Eq. (4.9) comes from the integral

$$\frac{i\omega}{4} \int \frac{q_3^2}{(q_1^2 + q_2^2 + q_3^2 - (q_0 - \frac{m}{2})^2 - \frac{\omega^2}{4})^2 + \omega^2 q_3^2} e^{-2(q_0^2 + \cdots + q_3^2)/g} dq_0 \cdots dq_3.$$

First of all, we perform the angular integration of the two-dimensional integral over $q_1$ and $q_2$. This gives, with the definition $q^2 = q_1^2 + q_2^2$

$$\frac{i\omega}{4} 2\pi \int \int_0^\infty \frac{q_3^2 q}{(q^2 + q_3^2 - (q_0 - \frac{m}{2})^2 - \frac{\omega^2}{4})^2 + \omega^2 q_3^2} e^{-2(q^2 + q_0^2 + q_3^2)/g} dq dq_0 dq_3. \quad \text{(B.3)}$$

Next we evaluate the $q$-integration for $\omega \sim 0$

$$\omega q_3^2 \int_0^\infty \frac{q e^{-2q^2/g}}{(q^2 + q_3^2 - (q_0 - m/2)^2 - \omega^2/4)^2 + \omega^2 q_3^2} dq$$
$$\sim \frac{\pi}{2} |q_3| e^{2(q_3^2 - (q_0 - m/2)^2 - \omega^2/4)/g} \Theta((q_0 - m/2)^2 + \omega^2/4 - q_3^2),$$

and carry out the $q_3$ integration

$$\int |q_3| \Theta((q_0 - m/2)^2 + \omega^2/4 - q_3^2) dq_3 = (q_0 - m/2)^2 + \omega^2/4.$$

Finally, the $q_0$-integration yields

$$\int [(q_0 - m/2)^2 + \omega^2/4] e^{-2(q_0^2 + (q_0 - m/2)^2)/g} dq_0 = \frac{\sqrt{\pi g}}{16} e^{-m^2/4g} (g + m^2/2 + 2\omega^2).$$

The combination all these results leads to an expression for Eq. (B.3) as

$$i\frac{\pi^2 \sqrt{\pi g}}{64} e^{-m^2/4g} (g + m^2/2 + 2\omega^2). \quad \text{(B.4)}$$

Figure Caption

Density of states as a function of the average mass $m$ at energy $E = 0$ and disorder strength $g = 2$. The $g/m^2$ expansion (full curve) is compared with the result of the large-$N$ limit (homogeneous saddle point solution) of Ref. [12] (broken curve).

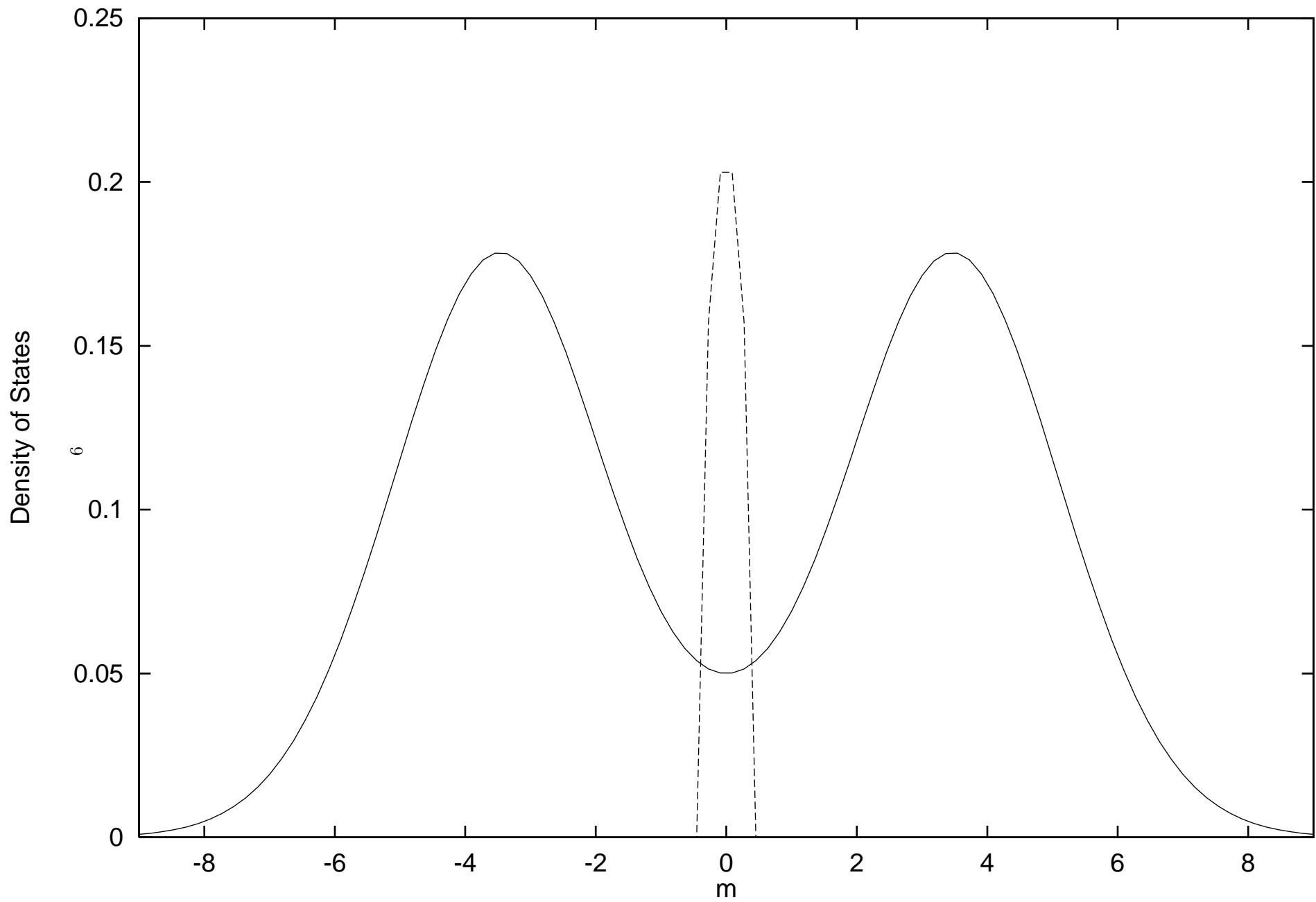